\documentclass[3p,times,procedia]{elsarticle}
\usepackage{nupha_ecrc}

\volume{00}

\firstpage{1}

\journalname{Nuclear Physics A}

\runauth{}

\jid{nupha}

\jnltitlelogo{Nuclear Physics A}

\usepackage{amssymb}

\usepackage[figuresright]{rotating}
\def\be{\begin{equation}}
\def\ee{\end{equation}}
\def\ba{\begin{eqnarray}}
\def\ea{\end{eqnarray}}

\begin{document}

\begin{frontmatter}



\dochead{XXVIIth International Conference on Ultrarelativistic Nucleus-Nucleus Collisions\\ (Quark Matter 2018)}

\title{Anisotropic hydrodynamic modeling of heavy-ion
collisions at LHC and RHIC}


\author[IAU]{Mubarak Alqahtani}
\author[Kent]{Dekrayat Almaalol}
\author[Kent]{Mohammad Nopoush} 
\author[Krakow]{Radoslaw Ryblewski}
\author[Kent]{and Michael Strickland}

\address[IAU]{Department of Basic Sciences, College of Education, Imam Abdulrahman Bin Faisal University, Dammam 34212, Saudi Arabia}
\address[Kent]{Department of Physics, Kent State University, Kent, Ohio 44242, USA}
\address[Krakow]{The H. Niewodnicza\'nski Institute of Nuclear Physics, Polish Academy of Sciences, PL-31342 Krak\'ow, Poland}

\begin{abstract}
In this proceedings contribution, we review 3+1d quasiparticle anisotropic hydrodynamics (aHydroQP). Then, we show some recent phenomenological comparisons between the aHydroQP model and some experimental  results. We show comparisons between aHydroQP and Pb-Pb 2.76 TeV collisions from the ALICE collaboration and Au-Au 200 GeV collisions from RHIC experiments. We show that the quasiparticle anisotropic hydrodynamics model is able to describe the experimental results for Pb-Pb and Au-Au collisions quite well for many observables such as the spectra, multiplicity, elliptic flow, and HBT radii in many centrality classes.

\end{abstract}

\begin{keyword}

Quark-gluon plasma, Relativistic heavy-ion collisions, Anisotropic hydrodynamics


\end{keyword}

\end{frontmatter}



\section{Introduction}
\label{}
Heavy-ion collision experiments at the Relativistic Heavy Ion Collider (RHIC) and Large Hadron Collider (LHC) create and study the quark-gluon plasma (QGP). From these experiments, it was concluded that the QGP is a strongly interacting system exhibiting clear collective behavior. This suggests using ideal, and later on, viscous relativistic hydrodynamics \cite{Huovinen:2001cy,Romatschke:2007mq,Ryu:2015vwa,Niemi:2011ix}. However, the QGP is a highly momentum anisotropic plasma at early times after the nuclear passthrough which motivates introducing anisotropic hydrodynamics \cite{Florkowski:2010cf,Martinez:2010sc,Martinez:2012tu,Ryblewski:2012rr,Bazow:2013ifa,Nopoush:2014pfa,Nopoush:2014qba}. Recently, the 3+1d quasiparticle anisotropic hydrodynamics model was introduced and compared to different heavy-ion observables. For recent reviews about anisotropic hydrodynamics, we refer the reader to \cite{Strickland:2014pga,Alqahtani:2017mhy}.

In this proceedings contribution, we will introduce anisotropic hydrodynamics and then focus on a specific approach, quasiparticle anisotropic hydrodynamics (aHydroQP) \cite{Alqahtani:2015qja,Alqahtani:2016rth}. Then, we will present comparisons between aHydroQP and some heavy-ion observables at two different energies. First, comparisons with Pb-Pb collisions at 2.76 TeV from the ALICE collaboration \cite{Alqahtani:2017jwl,Alqahtani:2017tnq,Alqahtani:2017gnn}. Second, comparisons with Au-Au collisions at 200 GeV from different RHIC experiments \cite{Almaalol:2018gjh}. 

\begin{figure*}[t!]
\centerline{
\includegraphics[width=.5\linewidth]{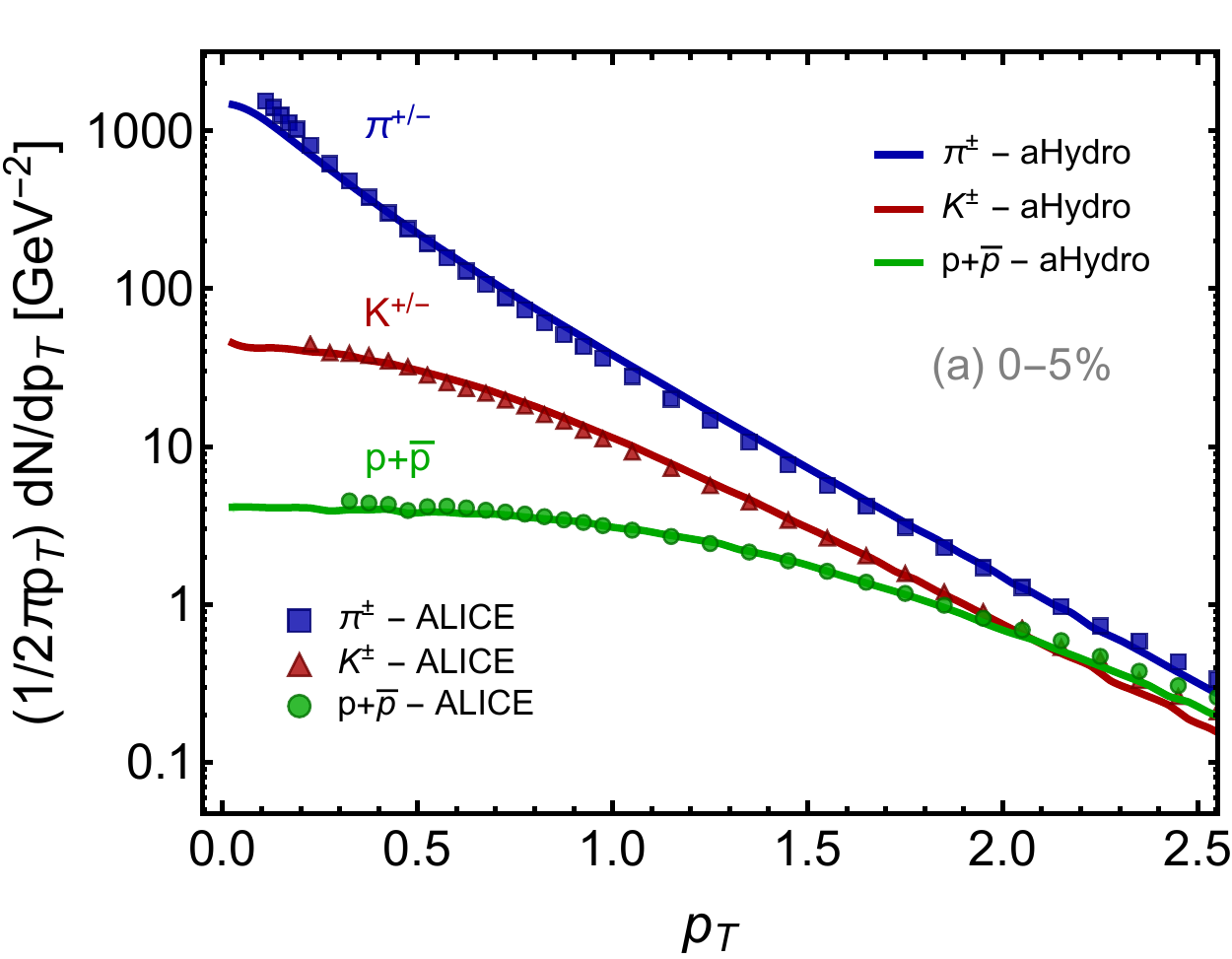}
\includegraphics[width=0.48\linewidth]{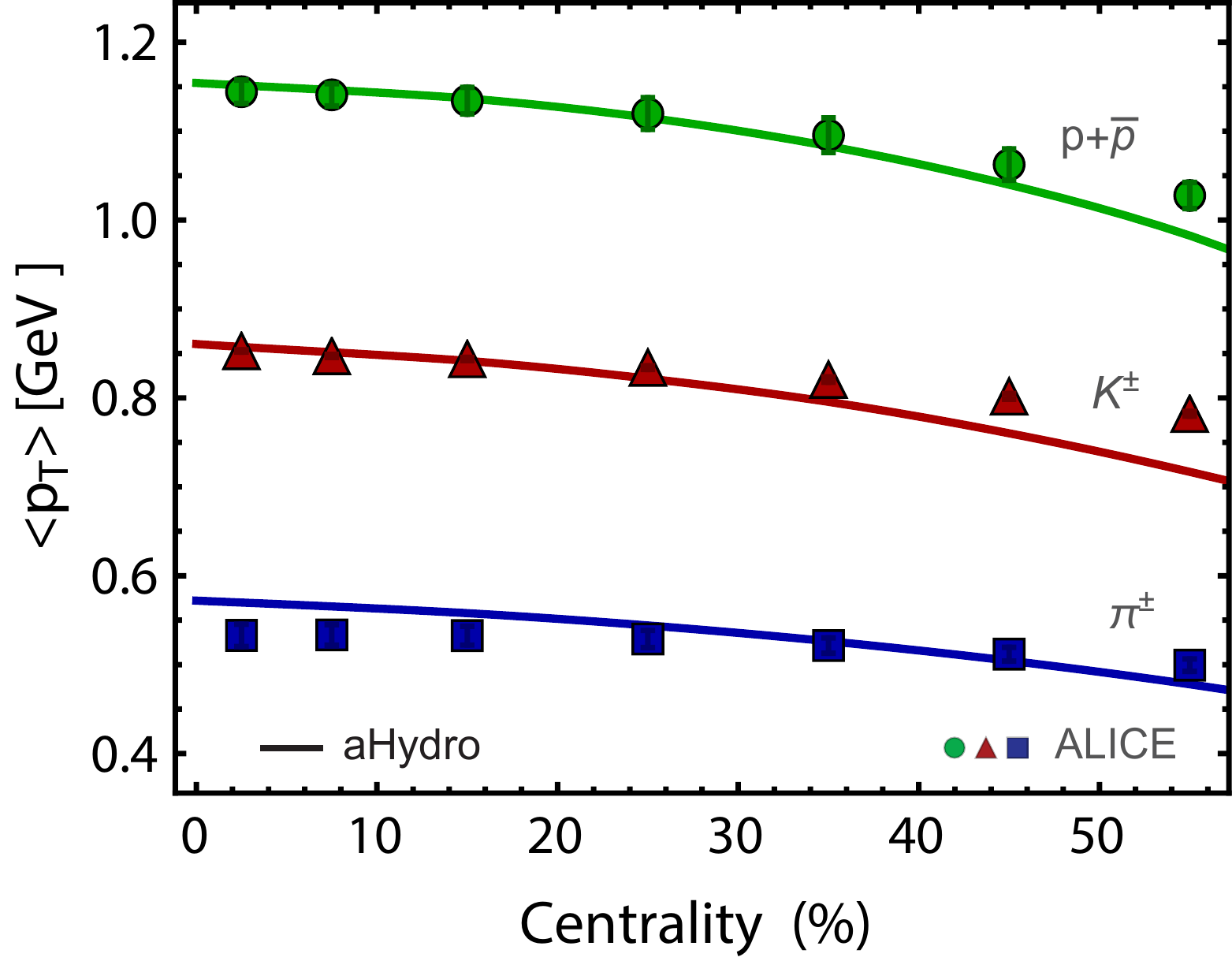}
}
\caption{In the left panel, the spectra of $\pi^\pm$, $K^\pm$, and $p+\bar{p}$ in 0-5\% centrality classes are shown where data are taken from \cite{Abelev:2013vea}. In the right panel, $\langle p_T \rangle $ of pions, kaons, and protons is shown as a function of centrality where data are also from the ALICE collaboration  Ref.~\cite{Abelev:2013vea}. Figure is taken from Ref.~\cite{Alqahtani:2017tnq}.}
\label{fig:ptavg}
\end{figure*}

\section{3+1d quasiparticle anisotropic hydrodynamics}
The dynamical equations for  3+1d  quasiparticle anisotropic hydrodynamics for massive relativistic  quasiparticle systems can be obtained by taking moments of the Boltzmann equation \cite{Alqahtani:2015qja}.
\be
p^\mu \partial_\mu f(x,p)+\frac{1}{2}\partial_i m^2\partial^i_{(p)} f(x,p)=-C[f(x,p)]\,,
\label{eq:boltzmanneq}
\ee
where the mass is a function of temperature and $C[f(x,p)]$ is the collisional kernel which is taken to be in RTA \cite{Alqahtani:2015qja}. In the local rest frame the distribution function is given by 
\be
f(x,p) =  f_{\rm eq}\!\left(\frac{1}{\lambda}\sqrt{\sum_i \frac{p_i^2}{\alpha_i^2} + m^2}\right) \, .
\label{eq:fform}
\ee
We note that by taking the anisotropy parameters $\alpha_x$ = $\alpha_y$ = $\alpha_z$ = 1 and $\lambda$ = T, one recovers the isotropic equilibrium distribution function \cite{Alqahtani:2015qja}.
\section{Phenomenological results}
\label{}

Now, we present comparisons between 3+1d quasiparticle anisotropic hydrodynamics and  experimental data from the ALICE collaboration for 2.76 TeV Pb+Pb collisions and 200 GeV Au-Au collisions. Due to the space limitation, we will present a small set of observables, and refer the reader for more comparisons and details to original references, see Refs.~\cite{Alqahtani:2017jwl,Alqahtani:2017tnq,Almaalol:2018gjh}. 

\begin{figure}[t!]
\centerline{
\hspace{-1.5mm}
\includegraphics[width=1\linewidth]{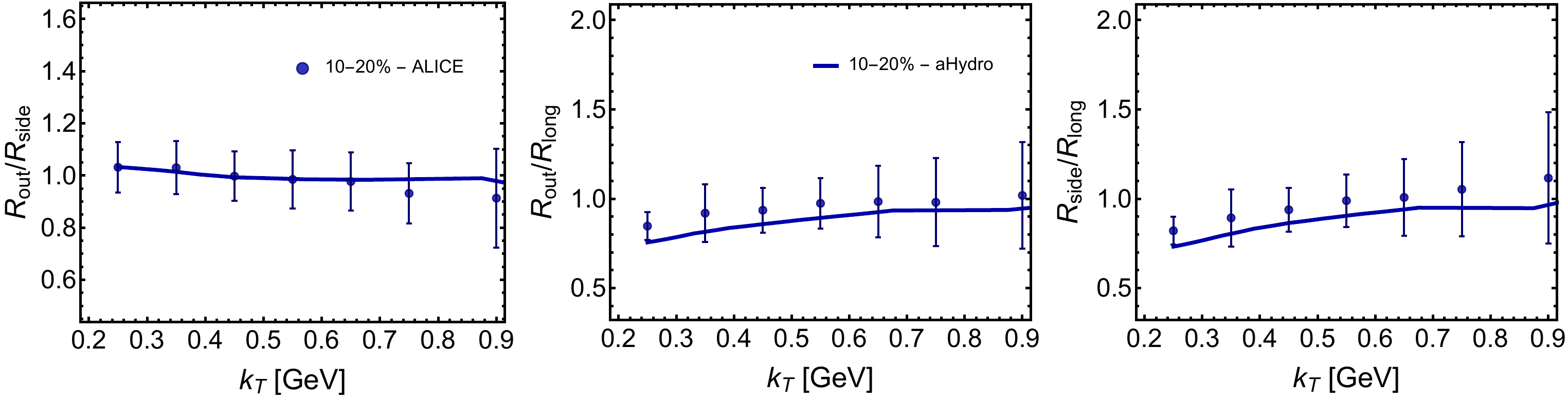}}
\caption{ The femtoscopic Hanbury-Brown-Twiss (HBT) radii ratios as a function of the pair mean transverse  momentum for $\pi^+ \pi^+ $ is shown for 10-20\% centrality class. The left, middle, right panels show $R_{\rm out} /R_{\rm side} $, $ R_{\rm out}/ R_{\rm long} $, and $R_{\rm side}/R_{\rm long} $, respectively.  All results are for 2.76 TeV Pb+Pb collisions where  data shown are  from the ALICE collaboration \cite{Graczykowski:2014hoa}. Figure is taken from Ref.~\cite{Alqahtani:2017tnq} }
\label{fig:HBT}
\end{figure}

First, let us start by showing the comparisons between aHydroQP and the experimental data from ALICE collaboration for 2.76 TeV Pb+Pb collisions. In Fig.~\ref{fig:ptavg}, in the left panel we show the spectra of pions, kaons, and protons as a function of the transverse momentum $p_T$  in 0-5\% centrality class. In the right panel, we show the average transverse momentum as a function of centrality. From both panels, we see that aHydroQP agrees with the data quite well. 
We also show in Fig.~\ref{fig:HBT}, comparisons of the HBT radii ratios $R_{\rm out}/R_{\rm side} $, $ R_{\rm out}/R_{\rm long}  $, and $R_{\rm side}/R_{\rm long} $, respectively in the 10-20\% centrality class, as a function of the pair average transverse momentum. As can be seen from this figure, aHydroQP predictions are in a good agreement with the experimental data.

\begin{figure*}[t!]
\centerline{
\includegraphics[width=.46\linewidth]{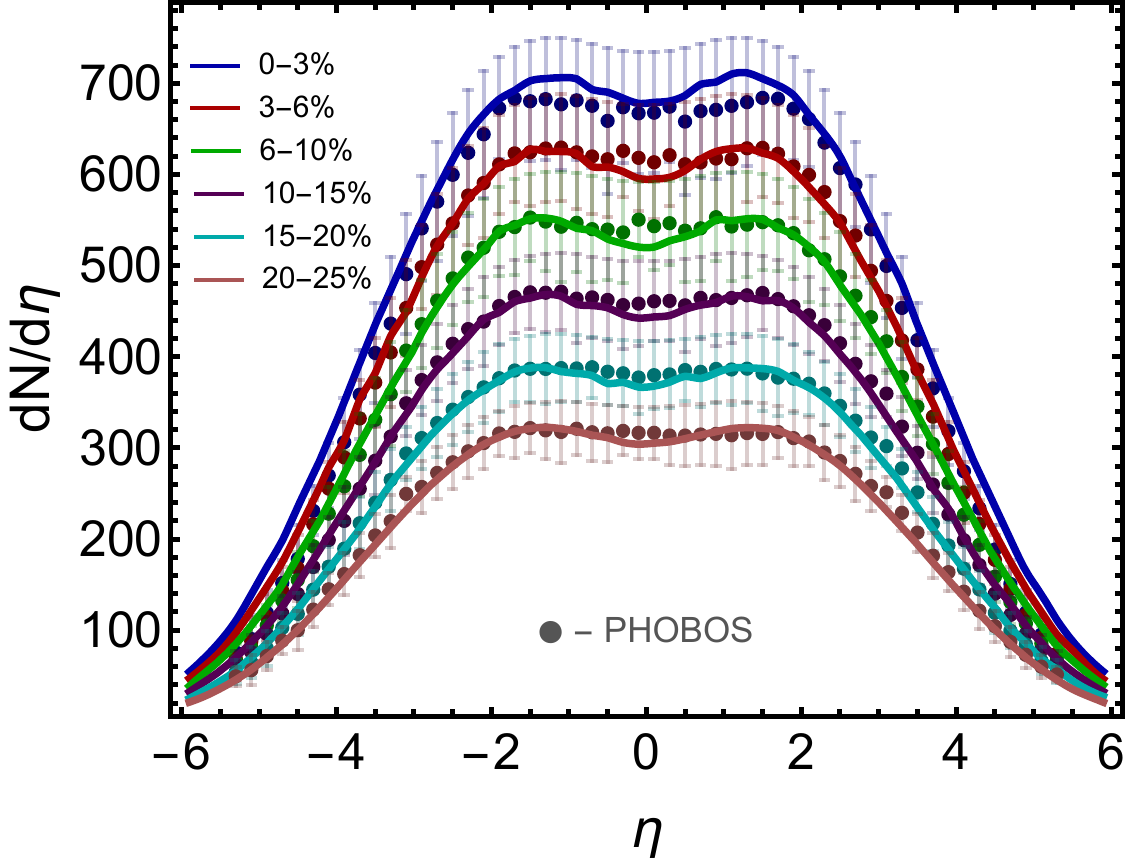}
\includegraphics[width=0.46\linewidth]{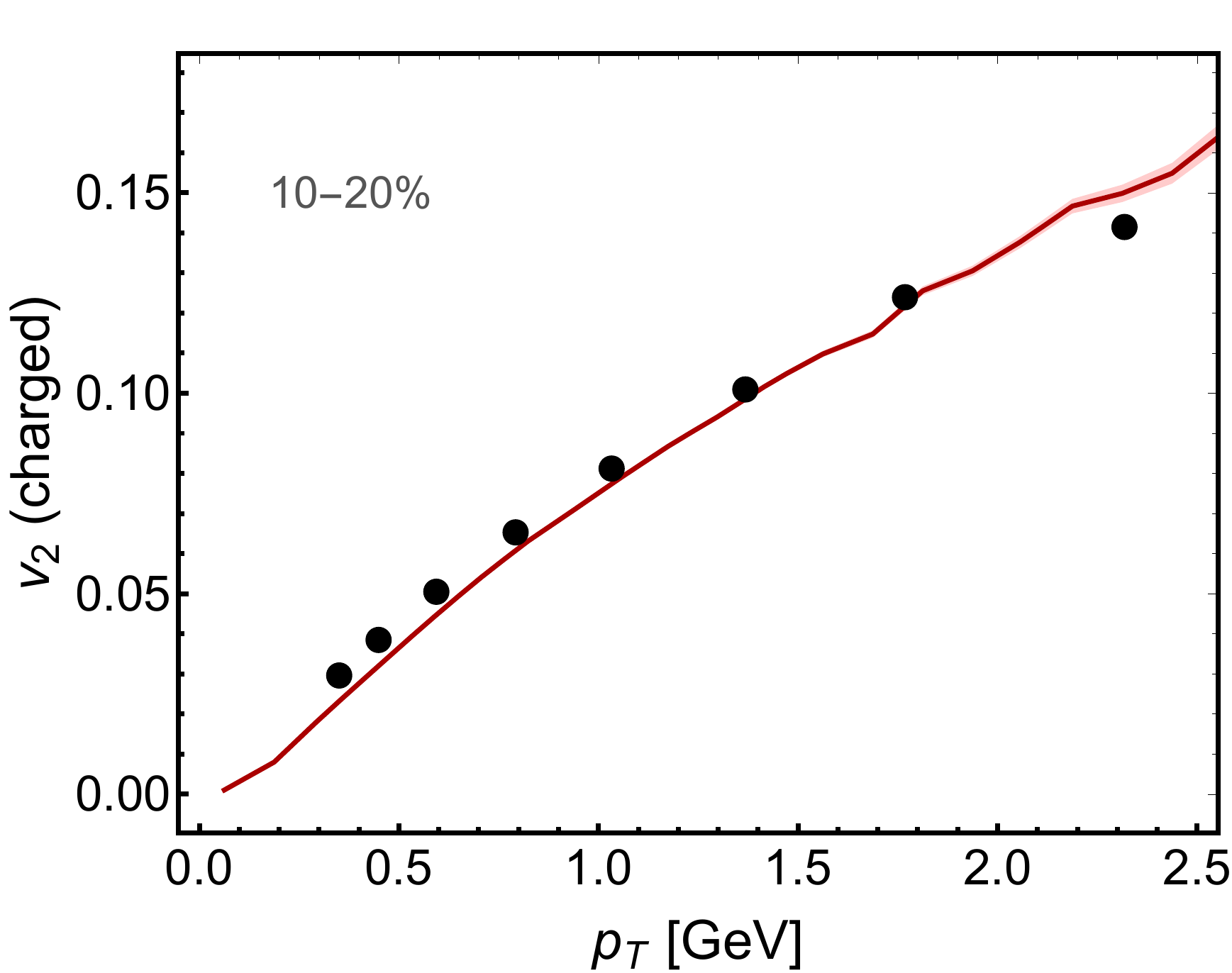}
}
\caption{In panel (a), the charged particle multiplicity in different centrality classes (0-25\%) is shown between aHydroQP and experimental data taken from the PHOBOS collaboration \cite{Alver:2010ck}. In panel (b), the elliptic flow for charged particles in 10-20\% centrality class is shown where data is taken from the PHENIX collaboration \cite{Adare:2006ti}. Figure is taken from \cite{Almaalol:2018gjh}.}
\label{fig:RHICresults}
\end{figure*}

Next, let us turn to comparisons between aHydroQP and experimental results at RHIC's highest energies. In Fig.~\ref{fig:RHICresults}-a, we show comparisons of the charged particle multiplicity as a function of pseudorapidity predicted by our model and experimental data from the PHOBOS collaboration \cite{Alver:2010ck}. We find the agreement between aHydroQP and the experimental results is quite good in a wide range of centrality classes. In Fig.~\ref{fig:RHICresults}-b, we present the elliptic flow for charged particles as a function of transverse momentum in the 10-20\% centrality class where the data is taken from the PHENIX collaboration \cite{Adare:2006ti}. As can be seen from this figure, our model shows good agreement with the experimental results.

Finally, we would like to list the extracted fitting parameters that we used in the above comparisons. In the ALICE case, the extracted parameters  are $T_0 (\tau_0=0.25$\,fm/c$)= 600$ MeV and $\eta/s = 0.159$. However, in comparisons at 200 GeV the extracted parameters  are $T_0 (\tau_0=0.25$\,fm/c$)= 455$ MeV, $\eta/s = 0.179$. The freeze-out temperature is fixed in both cases as \mbox{$T_{\rm FO} = 130$ MeV}.

\section{Conclusions and outlook}
In this proceedings contribution, we reviewed 3+1d quasiparticle anisotropic hydrodynamics. We next presented phenomenological comparisons with different heavy-ion collision experiments: Pb-Pb at 2.76 TeV and Au-Au at 200 GeV. Additionally, we listed our fitting parameters extracted from fits to the experimental  data in each case. Finally, we showed some observables like the spectra, the centrality dependence of the average transverse momentum, the elliptic flow as a function of the mean transverse momentum, and HBT radii. In conclusion, we showed that aHydroQP model was able to describe the experimental data quite well for many observables at different energies. 


\section{Acknowledgments}
M.~Alqahtani was supported by Imam Abdulrahman Bin Faisal University, Saudi Arabia.  D. Almaalol was supported by a fellowship from the University of Zawia, Libya. M.~Nopoush and M.~Strickland were supported by the U.S. Department of Energy, Office of Science, Office of Nuclear Physics under Award No. DE-SC0013470.  R. Ryblewski was supported by the Polish National Science Center grants No. DEC-2012/07/D/ST2/02125 and DEC-2016/23/B/ST2/00717.

\bibliographystyle{elsarticle-num}
\bibliography{aHydroQP}







\end{document}